\documentclass[showpacs,preprintnumbers,
amsmath,amssymb]{revtex4}
\usepackage{graphicx}
\usepackage{dcolumn}
\usepackage{bm}
\begin{document}
\title{ Extreme waves and modulational instability:
 wave flume experiments on irregular waves}
\author{M. Onorato$^a$,  A. R. Osborne$^a$, M. Serio
$^a$, L. Cavaleri$^b$,C. Brandini$^c$ and C.T. Stansberg$^d$ }
\affiliation{$^a$ Dip. di Fisica Generale, Universit\`a di 
Torino, Via P. Giuria, 1 - 10125 Torino, ITALY \\
$^b$ ISMAR 1364 S. Polo -
30125 Venezia, ITALY \\
$^c$ La.M.M.A., Regione Toscana, Via Madonna del Piano - 
50019 Sesto Fiorentino,  ITALY \\
$^d$ Norwegian Marine Technology Research Institute A.S
(MARINTEK), P.O. Box 4125 Valentinlyst,
N-7450 Trondheim, NORWAY}
\date{\today}
\begin{abstract}
We discuss the formation of large amplitude waves for
sea states characterized by JONSWAP spectra 
with random phases. In this context we discuss 
experimental results performed in 
one of the largest wave tank facilities 
in the world. We present 
experimental evidence that the tail
of the cumulative probability function 
of the wave heights for random waves strongly 
depends on the ratio between the
wave steepness and the spectral bandwidth. 
When this ratio, called the Benjamin-Feir
Index, is  large the Rayleigh 
distribution clearly underestimates the occurrence of
large amplitude waves.
Our experimental results are also 
successfully compared with previously performed
numerical simulations of the Dysthe equation.
\end{abstract}
\maketitle
\section{Introduction}
The determination of the probability density function 
of wave heights for a system of a large number
of random waves is definitely a task of major importance
both from  theoretical and applicative points of view.
For linear waves the fundamental work was
 done by \cite{RICE} in connection with noise in electronic
circuits. Some years later \cite{LONG} 
 adapted the ideas of Rice to surface gravity waves. He showed
that if the wave spectrum is narrow banded and if the
phases of the Fourier components of the surface elevation 
 are distributed uniformly 
(random phases), then the probability distribution
of crest-to-trough wave heights is given by the Rayleigh distribution.
In the study of extreme events,
it is useful to define the 
probability that the wave height $H$ assumes
a value greater than a reference height, $H_0$.
 This probability, also known as
the survival function $S(H>H_0)$,
is given by $1-P(H \le H_0)$, where $P$ is the cumulative 
probability function.
For the Rayleigh 
distribution the survival function
is given by: 
\begin{equation}
S(\xi > \xi_0)=Exp[-2 \xi_0^2], \label{survival}
\end{equation}
where $\xi=H/H_s$, with $H_s$ the significant wave height.
Using the survival function, it is for example 
possible to establish  
that the probability of finding 
a wave whose height is 
greater than two times the significant wave height
is about 1/2980. 
The value of two is usually selected as a 
threshold for identifying abnormally high waves (also
known as freak waves) in a  
surface elevation time series.

After the pioneering work by \cite{LONG},
the validity of the Rayleigh distribution for
wave heights has been widely investigated.
The distribution (\ref{survival}) was found 
to agree well with many field observations (\cite{EARLE75})
even though the frequency spectrum was not always
so narrow  and the steepness was not as small 
as required by the theory.
During the last 30 years, 
many empirical distribution functions have been proposed
to fit better the data.
\cite{FOR78} analyzed data recorded 
during hurricanes in the Gulf of Mexico and obtained
a better agreement with a two parameter Weibull distribution. 
According to his analysis, the 
Rayleigh distribution was over-predicting the 
experimental data.
Some years later \cite{LONG80} re-examined the same data and showed
that the Rayleigh distribution fitted equally well the data,
provided that the 
value of the root mean square of the amplitude is suitably
modified by introducing a finite 
spectral band width (see also \cite{NAESS85}).

In recent years particular attention has been given to 
understanding the mechanism of formation 
of freak waves and their influence
on the tail of the probability density function of
wave heights. 
Even though a number of physical mechanisms have been identified
(linear superposition, \cite{LONG}, the wave-current
 interaction, see \cite{LAV} and \cite{FORN}, and the
modulational instability, see \cite{HEN},
\cite{TRUL97},\cite{TUL}), it should be stated that
not much theoretical progress has been made concerning 
the resulting statistical properties of the surface elevation.
More in particular, the relation
between the various sea states and the probability 
density function has not been clearly identified. 

Limiting the study to one-dimensional 
propagation, \cite{ONO00} have performed numerical 
simulations of the Nonlinear Schroedinger and 
Dysthe equations with initial conditions
provided by the random JONSWAP spectrum with different values
of the enhancement factor $\gamma$ and the 
Phillips constant $\alpha$ ($\gamma$ is related to the 
spectral band-width and the wave steepness;
 $\alpha$ is responsible for
the energy content of the
surface elevation and therefore contributes to
 the wave steepness).
One interesting result obtained is that 
the probability density function 
drastically depends on these two parameters $\alpha$ and
$\gamma$. For
small values of $\alpha$ and $\gamma$ the 
Rayleigh distribution approximates
the data rather well,
but for large values of $\alpha$ and $\gamma$
the Rayleigh distribution clearly under estimates
the tail of the probability density function (see
Figure 6b in \cite{ONO00}).
For example, using the Dysthe equation, with a JONSWAP 
spectrum with $\gamma=6$ and
$\alpha=0.0081$ as initial condition,  the probability of 
recording a freak wave (defined as a wave
whose height is at least two times the corresponding 
significant wave height) is 1/630, 
almost 5 times greater than the 
one predicted by the Rayleigh distribution!
Strong departure from the Rayleigh 
distribution was also observed numerically 
by \cite{MORI00} and by \cite{BRA2001}, using 
the Higher Order Spectral method of 
\cite{YUE87}  for solving the 
Euler equations for surface gravity waves.
This departure from the Rayleigh distribution
 was attributed to the 
Benjamin-Feir instability mechanism
that, provided the spectrum is sufficiently narrow
and the steepness is sufficiently large,
 can take place also in random waves (\cite{ALB78}).

Following the ideas developed in \cite{ALB78} and 
successively in \cite{JAN91}, \cite{ONO03} studied
the instability of a narrow banded approximation of a 
JONSWAP spectrum, individuating the region of
instability of the spectrum in the $\alpha$-$\gamma$
plane. They found that from random
spectra, as a result of
the modulational instability,
 oscillating coherent structures may be excited
 (these structures are particular
``breather'' solutions of the NLS equation \cite{DYS99},
 \cite{OSB00}). 
 More recently \cite{JAN03} discovered that the 
region of instability predicted using the theory
developed in \cite{ALB78}
is not completely consistent  with direct simulations
of the NLS and Zakharov equations. He therefore developed a
kinetic equation that takes into account quasi-resonant
interactions and demonstrated good agreement between
theory and direct numerical simulations;
 moreover he was also able to compute from the 
theory some statistical properties of the surface elevation
such as for example the kurtosis. He found out that, 
if the ratio between the steepness and the spectral
bandwidth is large (the ratio is often referred to as 
the  
Benjamin-Feir Index (BFI), see also
\cite{ONO01}), the gaussian 
distribution underestimates the tails
of the probability density function for
the surface 
elevation. Note that in 
his analysis the statistics was 
computed only on free waves, the Stokes
contribution was not included.
 Even though it was not mentioned
in the paper, we expect that in such conditions also the
survival function for wave heights should be far from
being well described by (\ref{survival}).
One of the major results in \cite{JAN03},
that confirms previous results from  
numerical simulations
in \cite{ONO00} and \cite{ONO01}, is therefore
that the statistical 
properties of the surface elevation 
strongly depend on the BFI and therefore on the spectral
shape (see section \ref{sec:bfi}). This conclusion, 
limited to deep water waves and to
one dimensional propagation, has been reached 
from simplified models, without including any mechanism of 
dissipation such as wave breaking.

Concerning random wave experiments in wave tank facilities, 
in the past 15 years it has been recognized that 
at about 15-20 wave-lengths from the wave 
maker extreme individual
wave heights in random records may appear
more frequently than predicted by the Rayleigh distribution
(see for example \cite{STAN00} and references therein).
The increase of the kurtosis observed along the wave tank 
in \cite{STAN92} was interpreted  as a 
higher-order effect, attributed to the modulational instability.
Nevertheless, to the knowledge of the authors, even though 
JONSWAP spectra are the most common runs in wave flumes for
many different applications, 
there has not been any systematic experimental study devoted
to understanding the relation between the BFI
 and probability density function of
wave heights.

In this paper we discuss some interesting
experimental results that we have obtained in 
a large facility at Marintek, Trondheim (Norway).
Our main goal in this paper is to give some 
experimental support to the numerical and 
theoretical work  performed in recent years
that suggests the idea that the modulational instability
can be responsible for the formation of freak waves.
Our experimental results,
compared successfully with previously
performed numerical simulations of the Dysthe equation,
 show that the probability density 
function of wave heights depends strongly on the BFI.
The paper is organized as follows:  
Section \ref{sec:bfi} contains a derivation
of the BFI; with respect to 
previous work we extend the definition of the 
 BFI to arbitrary depth.
 Sections \ref{sec:exp1} and  \ref{sec:exp2} are devoted 
to the description of the experiment and of
the results. Discussions and conclusions are reported in 
section \ref{sec:con}.

\section{The Benjamin-Feir Index and its relation
to the JONSWAP spectrum}\label{sec:bfi}
\begin{figure}
\centerline{\includegraphics[width=0.5\textwidth]{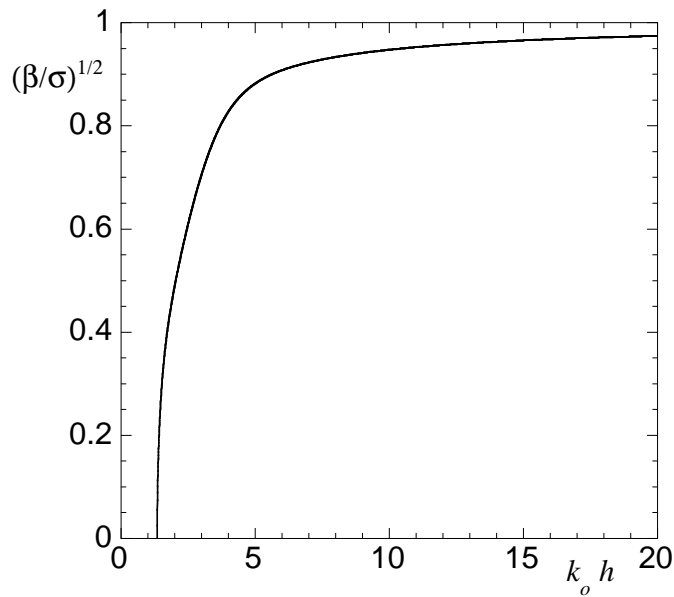}} 
\caption{\label{fig: bfi} Effect of finite water depth 
on the BFI}
\end{figure}

The Benjamin-Feir Index has been introduced formally 
in the paper by \cite{JAN03} and can be obtained
in two different ways. The first more laborious one
 consists in 
following the approach by \cite{ALB78}, i.e. one starts from 
the Nonlinear Schroedinger equation, derives a 
kinetic equation
for inhomogeneous surface elevation, 
performs a linear stability analysis of an
homogeneous random spectrum and
obtains its condition of stability
 (see \cite{JAN03} for details).
 The resulting condition simply states
that a spectrum is stable if the BFI is less than
one. A simpler approach, the one that will 
be presented here,
proposed in \cite{ONO01}, is based 
on dimensional arguments (note that in \cite{ONO01}
the square of the BFI was considered and was
 referred to as an Ursell number). 
Consider the dimensional 
NLS equation in arbitrary depth, 
in a frame of reference moving with the group velocity:
\begin{equation}
\frac{\partial A} {\partial t}+
i \sigma \frac{1} {8} 
\frac{\omega_0} {k_0^2} \frac{\partial^2 A} {\partial x^2}+
i \beta \frac{1} {2} 
\omega_0 k_0^2 |A|^2 A=0,
\end{equation} \label{NLS}
where $A$ is the complex wave envelope, $k_0$
is the carrier wave number corresponding and 
$\omega_0$ is the respective angular frequency.
$\sigma$ and $\beta$ are functions
 of the product $k_0 h$,
with $h$ the water depth, and both tend to 1 as 
$k_0 h \rightarrow \infty$. The analytical form
of  $\sigma$ and $\beta$ can be found for example 
in the book of \cite{MEI}.
The next step consists in adimensionalizing equation
(\ref{NLS}) in the following way: $A'=A/a_0$, $x'=x \Delta K$
and $t'=t (\Delta K/k_0)^2  \sigma \omega_0/8$,
where $\Delta K$ represent a typical 
 spectral bandwidth, $a_0$
a typical wave amplitude. Equation (\ref{NLS})
reduces to: 
\begin{equation}
\frac{\partial A} {\partial t}+
i \frac{\partial^2 A} {\partial x^2}+
i \bigg(\frac {2 \epsilon} {\Delta K/k_0}\bigg)^2
\frac{\beta} {\sigma} |A|^2 A=0,
\end{equation} \label{NLSA}
where primes have been omitted. $\epsilon=a_0 k_0$ is a measure
of  the wave steepness.
We define the Benjamin-Feir Index as
the square root of the coefficient that 
multiplies the nonlinear term (a factor of $2/\sqrt{2}$ is
included to recover the definition in \cite{ALB78}):
\begin{equation}
BFI= \frac{ {\sqrt{2}}\epsilon} {\Delta K/k_0} 
\sqrt{ \frac{\beta} {\sigma}}
\end{equation} \label{BFI}
The definition corresponds to the one 
in \cite{JAN03}, except 
for the term $\sqrt{{\beta}/{\sigma}}$ 
on the right hand side which we 
have included here to include the influence 
of the water depth. 
The effect of this last term on the BFI as 
a function of $k_0 h$ is 
shown in Figure \ref{fig: bfi}. As the water depth increases
the function tends to one and goes to zero
for shallower water. For values of $k_0 h$ smaller than about 
1.36, the BFI looses its meaning because,
as it is well known, the coefficient 
in front of the nonlinear term in the 
NLS changes sign and the equation becomes stable with 
respect to 
side band perturbations.
As the BFI index
increases the nonlinearity increases; therefore
we expect that the number of freak waves increases.
Note that this result was first obtained numerically
with numerical simulations of the NLS equation in \cite{ONO01}.

Normally one measures time series rather 
than space series; therefore, for
 the computation of the BFI from experimental data,
the term in the BFI for infinite 
water depth $\Delta K/k_0$ should be 
replaced by $2 \Delta f/f_0$ where $f_0$ 
is the frequency at the spectral peak
and  $\Delta f$ spectral-band width.
Therefore, given a time series,  we estimate the
BFI in the following way: 
the wave spectrum is computed; the spectral 
half-width at half maximum provides an estimate 
of the spectral-band width, $\Delta f$.
Methods such as those based on the quality factor
(see for example \cite{JAN91}) could also be used
for the calculation of the spectral width. Nevertheless 
for our purposes, we find our simple and 
straightforward method to be satisfactory.
Using the linear 
dispersion relation, the peak frequency is converted into 
the peak wave-number
 and the steepness is then computed as
$\epsilon=k_0 H_s/2$, where $H_s$ is the significant
wave height computed as 4 times the standard deviation
of the time series. 

\begin{figure}
\centerline{\includegraphics[width=0.5\textwidth]{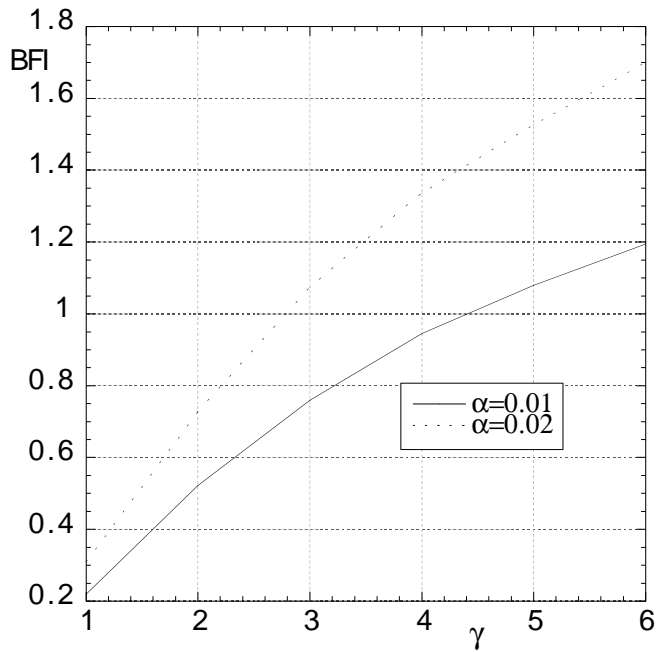}} 
\caption{\label{fig: bfijons} BFI as a function of the enhancement 
parameter $\gamma$ for $\alpha=0.01$ and  $\alpha=0.02$}
\end{figure}

We have applied this methodology 
in order to establish the relation between the BFI and 
the JONSWAP spectrum, see \cite{KOM}:
\begin{equation}
{P(f)= \frac {\alpha g^2}  {(2 \pi)^4 f^5} 
\exp \bigg[-\frac{5} {4} \bigg(\frac{f_0} {f}%
\bigg)^4\bigg]\gamma^ {\exp  [-\frac{(f-f_0)^2} {2\sigma_0^2f_0^2}]} 
,}  \label{jonswap}
\end{equation}
where $\sigma_0$=0.07 if $f\le f_0$ and $\sigma_0$=0.09 if $f> f_0$,
$g$ is gravity acceleration, $\alpha$ is the Phillips constant
and $\gamma$ is the enhancement parameter.
In Figure \ref{fig: bfijons} we show the BFI as a function of 
the $\gamma$ for  $\alpha=0.01$ and $\alpha=0.02$ (here
we have considered the case of infinite water depth. 
We note that for $\alpha$ fixed, larger values
of $\gamma$ imply a larger value of the Benjami-Feir-Index.
The Phillips constant $\alpha$ is strictly related to the 
wave energy, therefore to the wave steepness,
$\alpha \sim \epsilon^2$. If we double 
the value of $\alpha$ with $\gamma$ fixed the steepness
increase by a factor of $\sqrt{2}$ and so does the BFI 
(the spectral band-width remains practically unchanged.) 

\section{Description of the experiments} \label{sec:exp1}
\begin{figure}
\centerline{\includegraphics[width=1.\textwidth]{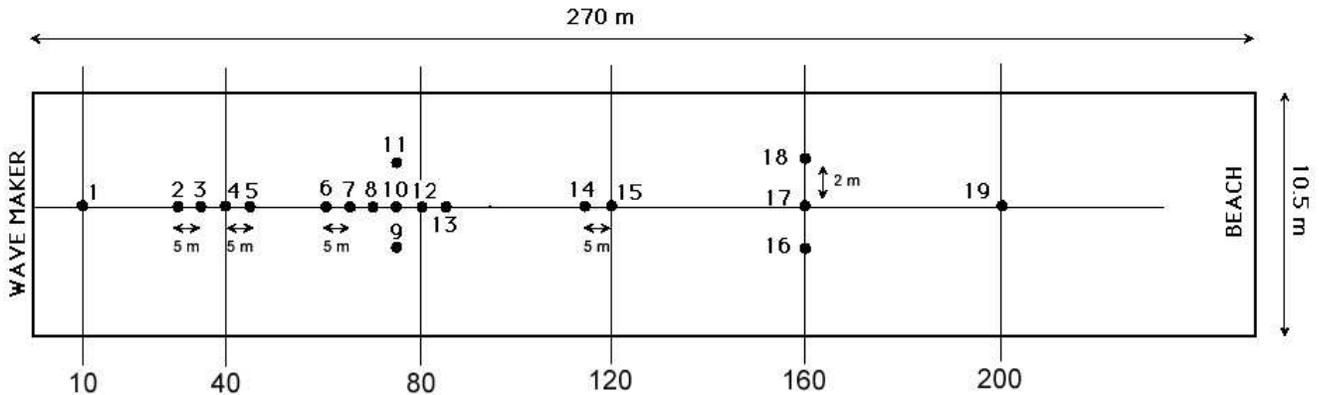}}
\caption{\label{fig: flume} Schematic of the wave tank facility at 
Marintek and location of wave probes.}
\end{figure}

The experiment was carried out in the long wave flume
at Marintek (see \cite{STANREP} for details).
 The length of the tank is $270$ $m$
and its width is $10.5$ $m$. The depth of the tank
is 10 meters for the first 85 meters and then
is reduced to 5 meters for the rest of the flume.
The effect of the jump
from 10 to 5 meters is insignificant for the
waves of 1.5 seconds considered here:
It can be easily seen from linear theory that 
the particle velocities at 5 meter depth are essentially zero.
A horizontally double-hinged flap type 
wave-maker 
located at one end of the tank was used to 
generate the waves. The distribution of signal 
frequencies to the upper and lower flap is 
automatically made by control 
software. All flap motion is computer controlled by using
pre-generated digital 
control signals stored in files.
A sloping beach is located at the far end of the tank
opposite the wave maker. After half an hour of an
irregular wave run  with peak period of 1.5 seconds, 
the wave reflection
was estimated to be 
less than 5\%.
The wave surface  elevation was measured 
simultaneously by
19 probes placed at different locations 
 along the flume (Figure \ref{fig: flume}).
Twin-wire conductance measuring probes were used; these
have excellent calibraion characteristics. Each wire was 
$0.3$ $mm$ in diameter, separated $10$ $mm$ from its twin
in the direction perpendicular to the main 
axis of the tank. 
A schematic of the flume with the location of the 
probes is shown in Figure \ref{fig: flume}. 
Preliminary simulations
with the one dimensional NLS equation were 
performed in order to estimate the 
spatial scales needed for the 
modulational instability to develop
in a random spectrum. Note that an estimation
of these scales 
based on the 
dispersion relation from small perturbation
theory of plane solutions
is not adequate because in a random
JONSWAP spectrum the perturbations are never
small! For steepness of around 0.1,  we estimated 
from numerical simulations that the instability would
take place at around
20-25 wavelengths, i.e for 1.5 seconds waves at 
around 60-75 meters from  the wave maker.
This is why a larger number
of probes were placed in that region.
 Lateral probes were also placed at 75 and 
160 meters in order to 
verify the quality of the long-crested waves 
generated at the wave maker.
The sampling frequency for each probe was 40 Hz.
JONSWAP random wave signals where synthesized 
as sums of independent harmonic components, by means
of the inverse Fast Fourier Transform of complex 
random Fourier amplitudes. These were
prepared according to the ``random
realization approach'' by using random spectral amplitudes 
as well as random phases. 
Three different JONSWAP spectra with different 
values of $\alpha$ and $\gamma$ have been investigated.
All of them were characterized by a peak period of 1.5 seconds.
In Table \ref{tab:parameters} we report the 
parameters that characterized each JONSWAP spectrum.
The three different experiments will be called 
BFI0.2, BFI0.9 and BFI1.2, with obvious meaning.
The value of $\sqrt{\beta/\sigma}$ in the BFI 
was estimated to be 0.95 at the wave maker.
\begin{table}
\begin{tabular}{ c c c c c}
$\gamma$ & $H_s (m)$ & $\epsilon=k_0 H_s/2$ 
& $\Delta f /f_0$  & BFI \\
\cline{1-5}
 1   & 0.11   & 0.098 & 0.28 & 0.2 \\ 
 3.3 & 0.14  & 0.125 & 0.09 &  0.9  \\
 6   & 0.16  & 0.142 & 0.08 &  1.2   \\
\end{tabular}
\caption{Parameters of the three different experiments performed
at Marintek}
\label {tab:parameters}
\end{table}
In order to have sufficiently good statistics,
a large number of waves was recorded.
Note that the large amount of data is of fundamental 
importance for the convergence of the tail 
of the probability density function for wave heights. Therefore
for each type of spectrum, 5 different realizations with 
different sets of random phases have been performed. 
The duration of each 
realization was 32 minutes. The total number of wave heights
(counting both up-crossing and down-crossing)
recorded for each spectral shape at each probe was  
about 12800 waves. 
In our analysis we have removed the first 
200 seconds of the records for each realization.
This lapse of time was calculated as the approximate time 
needed for the wave of frequency corresponding 
to twice the peak-frequency to
reach the last probe. 

\section{Experimental Results} \label{sec:exp2}

\begin{figure}
\centerline{\includegraphics[width=0.5\textwidth]{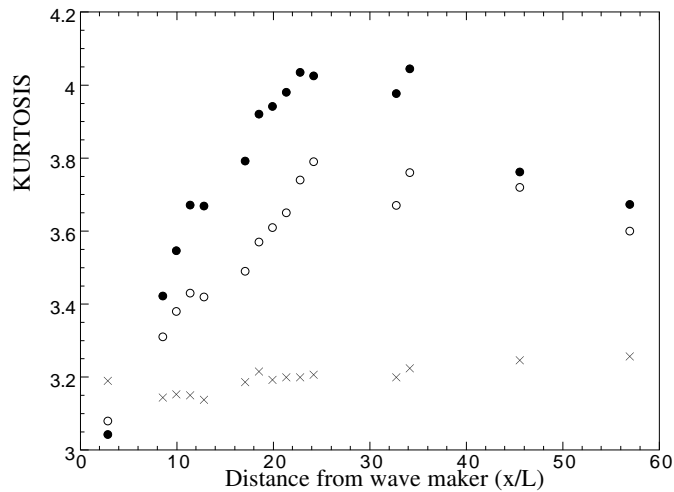}} 
\caption{\label{fig: kurt} Evolution of the kurtosis along the
wave tank: BFI0.2, crosses; BFI0.9, empty circles; BFI1.2 full 
circles (see Table \ref{tab:parameters}). 
The horizontal axis has been non-dimensionalized 
with the characteristic wavelength computed using the 
linear dispersion relation.}
\end{figure}

We first study the behavior of some statistical quantities
that can give an indication on
the presence of extreme 
events in the time series. In particular we consider the 
fourth-order moment of the probability density function, the
 kurtosis, that gives an indication of the importance of
the tail of the distribution function.
We recall that for a Gaussian distribution the
value of the kurtosis is 3, while larger values
of kurtosis in a measured
time series
can give an indication 
of the presence of extreme events. In Figure \ref{fig: kurt}
we show the kurtosis for the three experiments as
a function of the distance from the wave-maker.
The axes has been adimensionalized using 
the wave-length corresponding
to the peak period at the wave-maker: for $T$=1.5 seconds,
$L$=3.51 meters. First of all it should
be noted from the Figure 
that the kurtosis is always greater than the
 Gaussian prediction. 
For larger BFI (BFI=0.9 and BFI=1.2) the kurtosis 
grows very fast and reaches its maximum between 25 and
30 wave lengths from the wave maker. This result is
qualitatively consistent with our preliminary 
numerical analysis conducted with the NLS equation
from which it was estimated that the 
typical space scale of the modulational instability 
was of the order of 20-25 wave lengths. 
After 30 wave-lengths from the 
wave-maker the kurtosis decreases. This behavior 
could be due to a reduction of the modulational instability 
activity because of the reduction of the wave steepness:
waves have lost some energy due to 
wave-breaking (visible during experiments); moreover 
downshifting of the  peak period also takes place.
Locally, especially when the steepness is very large, 
three dimensional effects could also take
place \cite{TRUL99}.
For the smallest value of the BFI considered,
it is shown that the kurtosis is almost
constant with a mean value of 3.19.  This result
suggests a significant dependence on the
BFI of the statistical properties
of surface gravity waves. From Figure \ref{fig: kurt} 
we expect to find a larger number of extreme events
for larger values of the BFI. 
It should be mentioned that similar results as those
obtained for the BFI0.9 have been obtained previously
in \cite{STAN00}, carried out with approximately the same BFI, 
but in shorter and much wider basin (50 m x 80 m).
This indicates quite well that observed effects are independent
of the facility used in the experiment.

\begin{figure}
\centerline{\includegraphics[width=0.5\textwidth]{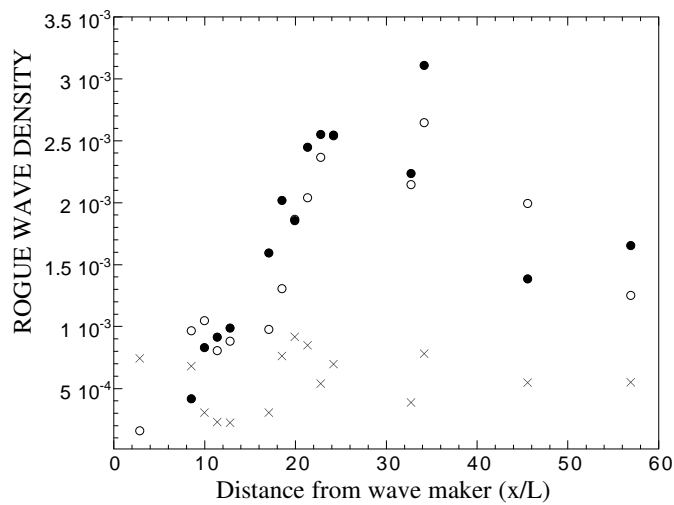}} 
\caption{\label{fig: rogdens} Rogue wave density for BFI0.2, 
crosses; BFI0.9, empty circles; BFI1.2, full circles
(see Table \ref{tab:parameters}).}
\end{figure}

We then turn our attention to wave heights and 
define from a time series 
the density of rogue waves as the number
of wave heights, considering both zero up-crossing plus
down-crossing waves, that satisfies
the conditions $H \ge 2 H_s$ over the total
number of wave heights recorded. In Figure 
\ref{fig: rogdens}
we show the rogue wave density for 
each probe at different distances from the wave maker.
Up to around 15 wavelengths from the 
wave maker the rogue wave density is bounded
between $10^{-4}$ and $10^{-3}$. While the density
 for BFI0.2 remains more or
less at the same level, the rogue wave density increases
substantially for  BFI0.9 and  BFI1.2 reaching a 
maximum of  $3.1$x$10^{-3}$ for BFI1.2.
Then in the last part of the tank the number of rogue waves 
decreases. This result is consistent
with the analysis of the kurtosis previously presented. 
According to our expectations,
the number of rogue waves recorded increases as
the Benjamin-Feir Index of the initial condition increases.

\begin{figure}
\centerline{\includegraphics[width=0.5\textwidth]{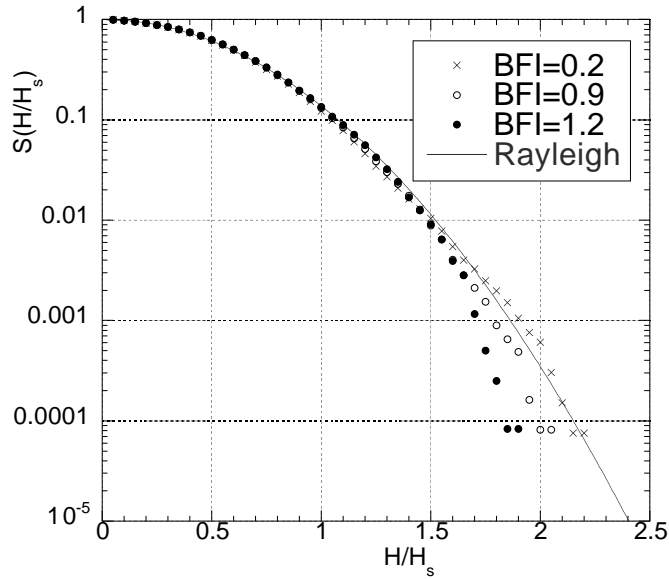}} 
\caption{\label{fig: surv1} Survival function at x/L=2.8 for  BFI0.2, 
crosses; BFI0.9, empty circles; BFI1.2 full circles.}
\end{figure}

We now discuss the behavior of the survival function for
wave heights, considering all together zero up-crossing and 
down-crossing wave heights. We compare our
experimental results with the
Rayleigh distribution (equation (\ref{survival})).
As previously stated the significant wave-height has
been computed as 4 times the standard deviation
of the time series. We compare the survival function
for the three different experiments at the same distance 
from the wave maker. In Figure \ref{fig: surv1}
we show the survival function at the first probe,
$x/L=2.8$. We recall that the wave field has been generated
at the wave maker
as a linear superposition of random waves; therefore,
we expect that at a few wavelengths the wave height 
should be described apprximately by the Rayleigh distribution.
For larger values of the BFI, the Rayleigh
distribution overestimates the experimental data for
large waves; this is consistent with most of the 
observation (see \cite{FOR78} and comments in \cite{LONG80}).
\begin{figure}
\centerline{\includegraphics[width=0.5\textwidth]{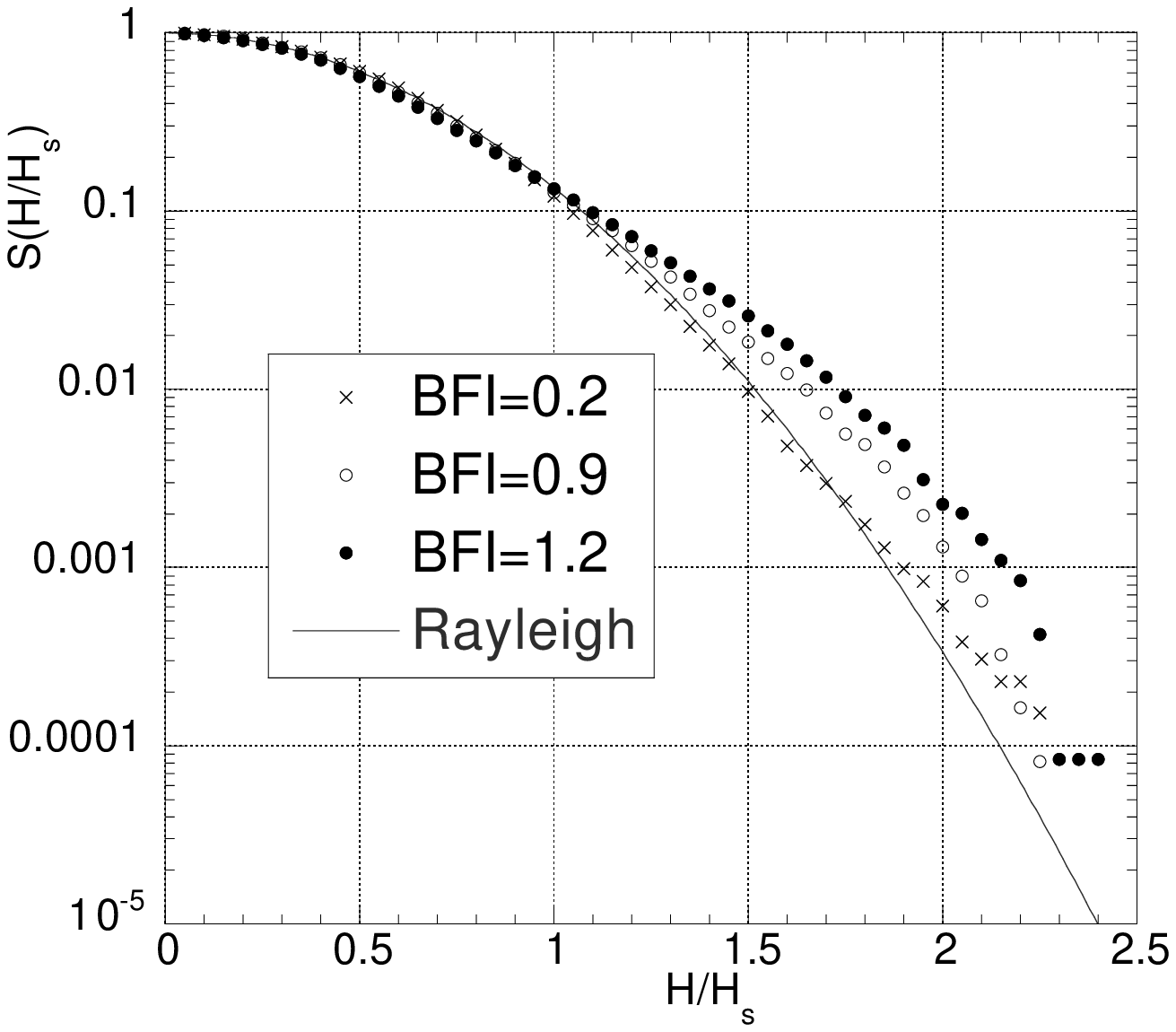}} 
\caption{\label{fig: surv2} Survival function at x/L=18.5 for  BFI0.2, 
crosses; BFI0.9, empty circles; BFI1.2 full circles.}
\end{figure}
\begin{figure}
\centerline{\includegraphics[width=0.5\textwidth]{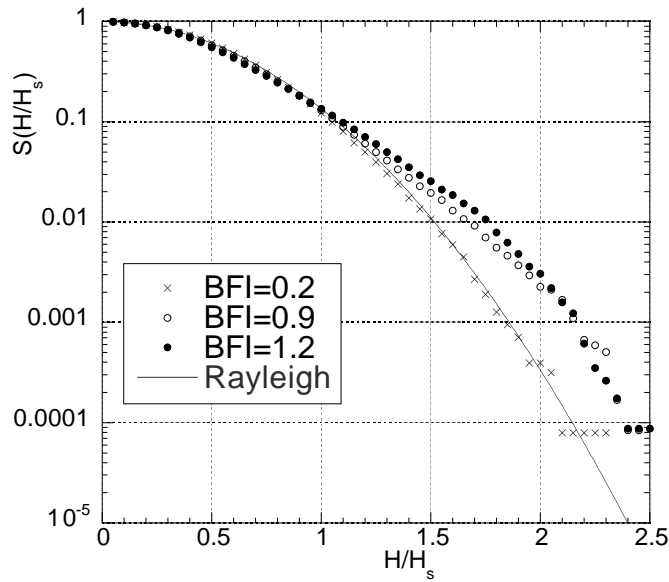}} 
\caption{\label{fig: surv3} Survival function at x/L=32.7 for BFI0.2, 
crosses; BFI0.9, empty circles; BFI1.2 full circles.}
\end{figure}
We then consider the probe at $x/L=18.5 $,
see  Figure \ref{fig: surv2}. While
the data from BFI=0.2 are well described 
by the Rayleigh distribution,
it is quite clear from the plot that the 
experimental data for BFI=0.9 and BFI=1.2 
are substantially underestimated by the Rayleigh distribution.
The curve for BFI=1.2 lies always on top of 
the one with BFI=0.9 and separates from the Rayleigh distribution
at around  $H/H_s$=1, which corresponds to a probability 
of 1/10 waves. A similar behavior is seen 
in Figure \ref{fig: surv3} at $x/L=32.7$.

\begin{figure}
\centerline{\includegraphics[width=0.5\textwidth]{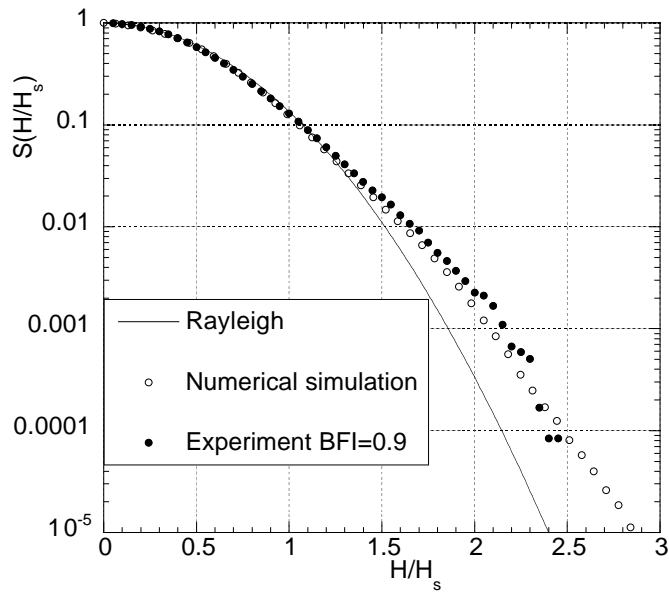}} 
\caption{\label{fig: surv4} Comparison of experimental results
with results from numerical simulation of the Dysthe equation, 
see \cite{ONO00} for details of the numerical
simulations.}
\end{figure}

We now compare in figure 
\ref{fig: surv4} our experimental results
with Figure 6b in \cite{ONO00} that
has been obtained by numerical simulations
of the Dysthe equation, using as initial condition a
JONSWAP spectrum with $\alpha=0.0081$ and $\gamma=6$.
The Benjamin-Feir Index calculated for this 
spectrum has approximately the value BFI=0.9,
therefore the probability distribution from
numerical simulations is 
compared with our experimental data with 
the same BFI. The agreement is quite good.


\section{Discussion and conclusions} \label{sec:con}

Recently a large number of papers that 
suggest that 
the modulational instability is a  possible 
mechanism for explaining the formation of 
 freak waves have appeared.
 A pioneering work in this context was performed by 
\cite{TRUL97}, who for the first time,
considered seriously the possibility
of describing nonlinear extreme waves using envelope 
equations. This work was also supported by a number of 
numerical and theoretical papers
in which particular analytical solutions (``breathers''
or more generally ``unstable mode'' solutions)
of the NLS equations were considered as
candidates for rogue waves (\cite{PER83},
\cite{DYS99}, \cite{OSB00}). An important
result that has been obtained 
from numerical simulations is that 
these ``breather'' solutions are very robust and can
be excited also from random spectra (\cite{ONO03}).
Their dynamics are quite different from the 
rest of the random wave field, they survive
after interactions with other waves
and periodically  appear as 
large amplitude waves, while disappearing after
some wavelengths. If the random wave field is
very nonlinear (large steepness and 
small spectral width, corresponding to 
large BFI), the density of these
objects increases, each of them having its own unique
dynamics. Starting from an initial random wave field, they 
need several wavelengths (in our case it was about 25-30 
wavelengths) to show up for the first time and, once
they are formed, they 
continuously appear and disappear, nonlinearly interacting with
the background wave field and each other. In a nonlinear 
wave field their presence is statistically 
significant, clearly leaving their 
signature in the survival function of wave heights.
This phenomenological description, that has been
conceived by us during many years of research
with the envelope equations (see \cite{ONO00},
\cite{ONO01}, \cite{ONO02}, \cite{OSB00}), is very consistent 
with the behaviour of real water waves. While for the NLS
equation those objects survive forever in 
an ideal simulation (indeed it can be shown 
that they correspond to proper modes, called 
unstable modes, 
of the NLS equation and therefore correspond to constants of the
 motion, as for example solitons can be considered as 
constant of motion for the Korteweg de Vries equation),
in real conditions higher-order effects can also take place:
 wave breaking and transverse instability 
can influence their dynamics
and transform an unstable mode to another 
less nonlinear unstable mode.
The experiments performed at Marintek
support this picture. Indeed we have shown that 
the number of rogue waves (which in the language
of the NLS equation roughly corresponds
to the number of unstable modes) depend on the Benjamin-
Feir Index of the initial spectrum. The presence of
these modes determines the shape of the 
probability density function of wave heights.
For nonlinearities consistent with water waves,
the density of these modes is not very large,
$3$x$10^{-3}$, therefore a large number of 
waves should be recorded in order to have reliable 
statistics.
Even if we have collected a large number of
data, sufficient to observe
a clear departure from the 
Rayleigh distribution, we believe that statistics 
computed  on even larger
data sets should be used in order
to model the tail of the survival function. 
The theory developed by \cite{JAN03} should
be closely compared with our experimental
results (this is already part of an ongoing research
program).

One of the limitations of our work 
that restrict us from extending our
results in a straightforward manner 
to real wind waves is that our experiment
has been performed in the case of infinite crested 
waves. For two dimensional propagation, numerical results and
experimental work on the probability density 
function of wave heights are
based on many fewer statistics, because numerics becomes much more
expensive and experimental work requires large basins with 
wave makers  capable of generating waves 
in different directions.
Only a few results are available:
\cite{ONO02}, using the Dysthe
equation in 2+1 dimensions with the exact linear 
dispersion relation, have
shown that, if directional spreading is sufficiently narrow, 
the kurtosis of the surface elevation reaches values 
larger than 3 (the value for Gaussian distribution). 
This result is consistent with experimental work 
carried out in the Ocean Basin at Marintek, \cite{STAN94}.
It should be stated that in 2+1 dimensions,
if the system is
not continuously forced, four wave resonant interactions
tend to broaden the spectrum  and
generate a tail of the form of $\omega^{-4}$ (\cite{DYS03});
therefore the instability may take place only at the 
initial stages of a freely decaying simulation.
Definitely more results including forcing and dissipation in 
2+1 dimensions are needed
for determining the role of the 
modulational instability for wind waves.

{\bf Acknowledgments}

We thank P. Janssen and K. Trulsen for
valuable discussions. Froydis Solaas is also 
acknowledged for technical support during the
experiment. This research has been supported by 
the Improving Human Potential - Transnational 
Access to Research Infrastructures Programme 
of the European Commission 
under the contract HPRI-CT-2001-00176.

\end{document}